# Hypersharp Resonant Capture of Neutrinos as a Laboratory Probe of the Planck length

R. S. Raghavan
*Institute of Particle, Nuclear & Astronomical Sciences and Department of Physics*
*Virginia Polytechnic Institute and State University, Blacksburg VA 24061*

The 18.6 keV antineutrino ( $\tilde{\nu}_e$) line from 2-body decay of $^3$H in crystals can be emitted with *natural* width because of motional averaging by lattice vibrations despite the very long lifetime of $^3$H and contrary to commonly held views of linewidths in such a case. It can be resonantly captured in $^3$He with geometrical cross section $\sigma \sim 10^{-17}$ cm$^2$. Using its hypersharp sensitivity $\Delta E/E \sim 10^{-29}$ and the method of "time-filtered" resonance, the basic energy width $\sim 10^{-24}$ eV expected of the $^3$H state can be measured to test if ultimate nuclear widths are limited by the Planck length rather than time-energy uncertainty.

Two-body nuclear weak decays emit monoenergetic lines of antineutrinos ($\tilde{\nu}_e$) in two well known modes: electron capture (EC) and bound-state beta decay (BB)[1] in which the β-electron is inserted in an atomic orbital instead of in the continuum. Recoilless emission (Mőssbauer effect ME) resulting in very precise energies of these lines was considered by Visscher for the EC mode[2] and Kells and Schiffer[3] for the BB decay of tritium $^3$H (T). However, these very early ideas remain speculative because of the very stringent, unanswered experimental demands. Current H-storage technology and materials now suggest a breakthrough in the T case. In a preliminary report[4], I proposed a specific approach to observe recoilless resonant capture of the 18.6 keV $\tilde{\nu}_e$ emitted in the T-BB in a $^3$He target. The key idea focuses on solving the biggest problem in this experiment—the difference of the noble gas He absorber and the chemically bound source T in metals.

With the advantage of recoilless emission, the resonant cross section σ for $\tilde{\nu}_e$ capture is determined by the spectral widths of the emitted $\tilde{\nu}_e$ which could be limited by line broadening induced by various means. A typical cause is the energy fluctuation due to dipolar interactions. In ref. 4, a relaxation width measured by NMR in the chosen material was used and a resonance cross section $\sigma \sim 3\times10^{-33}$ cm$^2$, very large for a $\tilde{\nu}_e$ reaction, was derived. While this was very attractive, major experimental challenges remained.

These prospects are now vastly improved by new ideas in this paper on the origins of the line-width for long lived states which basically remove the challenges faced in ref. 4. The broadening assumed in ref. 4 is appropriate for short lived γ-resonance states (all cases observed so far) but not for very long lived states. Inspired by well known results from other contexts of line narrowing, we show here that one should indeed expect hypersharp $\tilde{\nu}_e$ lines of *natural line width* because of the key role of motional averaging via harmonic lattice vibrations that was ignored in ref. 4. In this case σ can rise to the geometrical limit, $10^{-17}$ cm$^2$, vastly larger than the estimate in ref. 4 and enhance prospects for observing resonant capture of tritium $\tilde{\nu}_e$. In this paper I summarize the new ideas on $\tilde{\nu}_e$ linewidths and the simplified experimental approach.

A hypersharp T-$\tilde{\nu}_e$ line would be an extraordinarily powerful tool, combining a) high resonance cross section $\sigma \sim 10^{-17}$ cm$^2$; b) low $\tilde{\nu}_e$ energy (<20 keV); and c) hypersharp energy sensitivity $\Delta E/E \sim 10^{-29}$ that can probe new perspectives of the physical universe. Feature a) enables the overall feasibility of the T $\tilde{\nu}_e$ resonance experiment in practice. Feature b) can lead to a test of $\tilde{\nu}_e$ $\theta_{13}$ oscillations[4] in bench-scale—not km scale—baselines with grams-- not kilotons-- of target material; and c) opens the entirely new prospect of probing the fundamental quantum mechanical time-energy uncertainty in the untested regime of extremely small energies. A breakdown of this relation may result from upper limits on nuclear level widths imposed by a fundamental or Planck length $\mathcal{L} \sim 10^{-33}$ cm, in the regime of energy widths probed by the tritium $\tilde{\nu}_e$ resonance.

The BB decay $^3$H(1/2)$^+ \rightarrow$ $^3$He(1/2)$^+$ + $\tilde{\nu}_e$ ; E( $\tilde{\nu}_e$ ) = 18.6 keV; $\tau$ ($^3$H) ~ $6\times10^8$ s; $\Gamma \sim 10^{-24}$ eV, is ideal for resonant $\tilde{\nu}_e$ capture. It offers a sizable BB branching (~5.4 $\times10^{-3}$)[1] to the atomic ground state of $^3$He. The initial T atom has a vacancy in the 1s shell and the target $^3$He has two 1s electrons one of which can be captured in the reverse absorption process.

In BB decay,[1] $A(Z\text{-}1) \rightarrow \tilde{\nu}_e + A(Z) + e^-$ *(bound),* the $\tilde{\nu}_e$ line is emitted with the energy $E_{\tilde{\nu}e} = Q + B_Z - E_R$. The shell binding energy $B_z$ is *gained* in inserting an electron in A(Z). $E_R$ is the deficit due to nuclear recoil. Q is the maximum $\tilde{\nu}_e$ energy (= $M_{Z-1}$–$M_Z$) in the β-decay A(Z-1) → A(Z). In the absorption, $\tilde{\nu}_e + A(Z) + e^-$ *(bound)* $\rightarrow A(Z\text{-}1)$, $E(\tilde{\nu}_{e\ res}) = Q - B_Z - E_R$. The $\tilde{\nu}_e$ from the BB decay has exactly the excess energy $B_z$ for removing the same electron in the capture. The latter is resonant if $E_R$=0. i.e., $\tilde{\nu}_e$ emission and absorption are recoilless [3,4]

The resonance cross section σ is determined by the spectral density of the incident $\tilde{\nu}_e$ beam at the resonance energy. Broader the line, the smaller is the σ. A line width $\sim 10^{12}$ times the natural width was assumed in ref. 4. As discussed here, motional averaging creates $\tilde{\nu}_e$ of *natural* width Γ for long lived states such as T. Then the spectral density at resonance is maximal, 1/Γ. Thus, σ becomes maximal-- the geometrical value $\sigma_o = 2\pi^2 ƛ^2 \sim 2x10^{-17}$ cm$^2$, unprecedented for a $\nu_e$ reaction.

In a simple crystal, e.g. a metal, atomic motion is controlled by lattice vibrations. The nuclei sense a dipolar field H~ $\mu_1.\mu_2 / r^3$ (**r** is the interatomic distance), typically 1 gauss. This (inhomogeneous) field is *not static* because **r** (thus the field and the line energy) fluctuates via lattice vibrations. The key idea here is that lattice vibrations can motionally average out the dipolar field so that the line energy is hypersharp as seen in a simple estimate. For T in a Nb lattice (used below), field fluctuations $\Delta H$ ~0.02 gauss of T dipoles ($\mu$ = 3 $\mu_N$; I=1/2; $\gamma_N = 6\ \mu_N/\hbar$ ) occur due to mean vibrational displacement ~0.15A (indicated by the ME recoilless fraction) in a time $\tau_{latt}$ ($\sim\hbar$ /0.1 eV zero point energy). They imply a spin relaxation time[5] $T_2 \sim [(\gamma_N\ \Delta H)^2 \tau_{latt}]^{-1}$ $\sim 4x10^8$ s $\sim \tau = 5x10^8$ s, the T lifetime. Thus, the dipolar interaction in *vibrating* lattices does not preclude $\tilde{\nu}_e$ line emission from T- BB decay with natural width.

It was shown long ago that an *external* oscillating field can narrow dipolar broadening of the Mőssbauer (ME) line from long lived states[6]. The effect of harmonic fluctuations on shapes of ME lines including the role of the *nuclear* life time is explicitly shown in a frequency modulation (FM) approach to the emitted line shape by Salkola and Stenholm[7]. They used a simple modulation model $\Omega_o cos\Omega t$ due to an external oscillating field. Harmonic lattice vibrational motions create oscillating local fields via changes in r rather than by the $\mu$H interaction. The modulation model is a template for vibrational energy fluctuations and contains the essential features of the problem. In this model, $\hbar\ \Omega_o$ is the energy spread (from external fields or due to atomic vibrational displacements) and $\Omega$ = ($\hbar$ /T$_R$ = fluctuation time). The line shape is [7]:

$$A \propto \frac{1}{\Gamma} \Sigma_{k=-\infty}^{k=+\infty} J_k^{\,2}(\eta) \frac{1}{[(\delta/\Gamma)-k\xi]^2+1} \qquad (1),$$

where $J_k(x)$ are Bessel functions, $\eta = \Omega_o/\Omega$ , $\xi = \Omega/\Gamma$ and $\delta$ is the external detuning for scanning the line shape. The central line is obtained with $\delta$ ~0. The signal consists of a central line and an infinite series of sidebands of index ±k, all with the natural width.

The broadening of the central line is due to the overlap of the first sidebands with k= ±1 with the k=0 term in (1). The sidebands occur at energies $\delta$ tuned to make the square bracket in the denominator in (1) zero. The first side band (k=1) thus occurs shifted from the central line by $\delta/\Gamma = \xi$ *linewidths*. Thus, larger the $\xi = \Omega/\Gamma$, less the overlap, even though the absolute value of $\delta$ eV of the k=1 sideband position is set only by $\Omega$. $\xi$ increases *naturally* in long lived states as *Γ decreases*. A narrow central line of natural line width, well resolved from the side bands is thus achieved naturally and *necessarily* in the case of long lifetimes. In contrast to long lived states, short lived states (small $\xi$) emit a wide central line as well as sidebands. They are thus poorly resolved and with $\xi = \Omega/\Gamma \sim 1$ or less, line broadening occurs. As $\Gamma \to 0$, $\xi = \Omega/\Gamma >> 1$, thus hypersharp lines arise naturally from (and only from) long lived states contrary to conventional wisdom.

These ideas are closely analogous to the mechanism of the ME itself as shown by Shapiro[8] in a FM approach to ME line shapes subject to fluctuating Doppler shifts via nuclear motion in a vibrating lattice. The time-dependent displacement x(t) can be expanded in a series of the vibration frequencies $\Omega_m$. Then the wave field

$$E = \prod_m \Sigma_n\ J_n\left(\frac{x_m}{ƛ}\right) \exp[i(\omega_o + n\,\Omega_m)t].$$

The (recoilless) fraction of the unshifted line with n = 0 is given by[8] $f = \prod_m J_o^{\,2}\ (x_m/ƛ)$. Since $x_m/ƛ = x_m\omega_o/c$ <1, and $J_o(y) \approx 1-(y^2/4)$, we get the well known expression $f = 1- \Sigma(x_m/ƛ)^2 = \exp-[\langle x\rangle^2 ƛ^2]$ where $\langle x\rangle^2 = ½\ \Sigma(x_m)^2$. The ME line is accompanied by sidebands at $\omega_o \pm n\Omega_m$ each of intensity $J_n$. The line width depends on the overlap from the first sidebands. The side band resolution requires a *lifetime* $\tau >> 1/\Omega_m$ [9] satisfied in all ME cases studied so far. With the basic similarity of the ME and hypersharp line effects typified by (1), we can define a generalized hypersharp fraction $\mathcal{H} = J_o^{\,2}(\langle x\rangle/ƛ)\, \Pi_K\, J_o^{\,2}(\Delta_K/\Omega_K)$ where K runs over the different types of fluctuations with width $\Delta_K$ and rate $\Omega_K$. The recoilless fraction f is now just one of factors that determine the hypersharp line intensity. In summary, the central idea in this paper is harmonic averaging of all r-dependent energy fluctuations, e.g., lattice excitations, not only dipolar interactions.

Hypersharp $\tilde{\nu}_e$ emission in TBB→$^3$He based on lattice vibrational averaging, requires T and $^3$He (normally gases) to be embedded in *solids.* Metal tritides[10] offer a practical approach. T reacts with metals to form tritides and creates a uniform population of T in the bulk of the metal. As the tritide ages, the $^3$He daughter grows and uniformly populates the lattice ("the tritium-trick" TT). The He site in the source is its birth site –that of its parent T. In bcc metals such as Nb (of particular interest here), the T sits only in tetrahedral interstitial sites (TIS). The absorber is made in an identical TT method. However, the absorber site of He, an insoluble mobile inert atom, is typically different and indeed, non-unique. $^3$He tends to rapidly diffuse away and forms clusters or micro-bubbles, very different from the well defined sites of T and thus unsuitable for $\tilde{\nu}_e$ resonance.

The key experimental design problem is the search for a metal system where He sits at *well defined sites*

identical to T sites. A search was made using measured parameters from state-of-the-art tritide research such as

*Table 1 He transport parameters in NbT at 200K*

| $M_1T_1$ | E1 eV | E2 eV | E3 eV | D/cm$^2$ s |
|---|---|---|---|---|
| M=Nb | 0.9[a] | 0.13[b] | 0.43[b] | 1.1E-26[c] |

[a] Ref. 10; [b]Ref. 11; [c] With *tritium* pre-exponential $D_0$ (ref. 10)

*Table 2. Theoretical lattice energy data for T and $^3$He in Nb interstitial sites (IS) (Ref.12)*

| Site | EST (eV) | | ZPE (eV) | |
|---|---|---|---|---|
| | T | He | T | He |
| TIS | -0.133 | -0.906 | 0.071 | 0.093 |
| OIS | -0.113 | -0.903 | 0.063 | 0.082 |

1) He diffusivity D(K) at temperature K, 2) the He generation rate g =(T/Metal M)x1.79x10$^{-9}$/s, and 3) the activation energies E1, E2 and E3 for jumps, pair cluster formation and bubble coalescence[11] (see Table 1). A set of coupled non-linear differential equations[11] describe the time evolution of the concentrations c1 (interstitials IS), c2 (pair clusters) and c3 (bubbles). These equations were solved numerically for a variety of tritides, to observe the growth of He for 200 days at which time the T is switched off by desorption. Thereafter, the He in the T-free sample has different ratios of [interstitial sites IS/(bubbles)]=c1/(c2+c3) for different loading temperatures. At 200K, the He IS population grows linearly and remains indefinitely without bubble formation after the T is removed. For T>235K, c1 has a decaying profile indicating growing loss to bubbles. Thus, in NbT if T is maintained at <200K the $^3$He reside indefinitely *only* at unique IS sites. This behavior in NbT is exceptional (in PdT, bubble formation dominates already at >20K).

In the bcc NbT source, the T and thus, also the just-born daughter He reside in the TIS[12] (see Table 2 for the self-trapping energy EST at TIS and octahedral IS (OIS) sites. The EST at the two sites is degenerate. Thus, both sites can be randomly filled with equal probability. The site occupations have been verified by ion channeling.[13] With 6 TIS and 3 OIS in the bcc unit cell He sits at TIS identical to that of the emitter T-He 67% of the time. The NbT system thus uniquely meets the stringent demands of T-He matrix viable for resonant $\tilde{\nu}_e$ capture.

In the IST sites in Nb both the T and the He reside in deep potential wells (depths EST see Table 2). In the decay T→He the nearest neighbors are displaced by a small isotropic *local* dilation of the cell that is exactly reversible in He→T. The local dilation is a coherent antiphase isotropic *translational* motion of several Nb atoms that come to rest by the opposing force of the crystal which thus takes up the total momentum. Any lattice excitations due to the dilation occur only with the speed of sound, thus, long after $\tilde{\nu}_e$ emission. Thus they do not affect the $\tilde{\nu}_e$ energy.

When T and He in the TIS reside in potential wells they are isolated and inhibited from interactions with the lattice as well as random defects in it. The resonant atoms are harmonic oscillators in a *local* box. The only excitations of the recoiling atoms are to vibrational states $E_i$ *in the well*[14]. The $E_i$ in NbT have been calculated[12] and *measured* by neutron inelastic scattering[15] as $E_i$ =72 and 101 meV (doublet). Then f(T)= exp-[$E_R$(=62 meV) ($\Sigma 1/E_i$)] =0.125. The $E_i$ for He are unknown but using T data for He, f = f(T)f(He) ~1.5%.

In general, the $\tilde{\nu}_e$ energy is modified by energy *shifts* $E_T$ and $E_{He}$ (e.g., atomic shell energies of electrons and binding in local potential wells), the zero point energy (ZPE) in the local wells and the second order Doppler effect (which can be zeroed by identical temperatures for source/absorber)). The averaging effect due to harmonic lattice vibrations applies to all r dependent energies such as the local well depths and ZPE's which are averaged to unique *central* values. Thus the shifts $\Delta(E_T \pm E_{He})$ also attain *hypersharp values.* For example, in the decay T↔He there could be random misfits in the traps resulting in spreads of the net shift by $\Delta(E_T \pm E_{He})$. The Δ is averaged by the ZPE motion in the trap. For a spread Δ ~1%, the parameter $\eta = \Omega_o / \Omega$ in eq. (1) is Δ/ZPE ~ 8/90 = ~0.1 (using Table 2). By eq. 1, A = $J_o^4(0.1)$ = 0.99. Thus $(E_T \pm E_{He})$ will be practically hypersharp and exactly canceled in the emission ↔ absorption process. Note also that all multipole moments higher than dipole are zero for the spin ½ of $^3$H and $^3$He. Thus inhomogeneous broadening from electric fields due to random lattice defects are absent.

Practical geometries must be carefully designed to solve the problem of the gravitational red shift 10$^{-18}$ eV/cm >>Γ. The $\tilde{\nu}_e$ energies which depend on the gravitational potentials of the nuclei distributed in the source/absorber can be averaged to a unique central value by slow vertical oscillations. In principle, the central values in the source and absorber may not coincide. The resonance could be restored by an additional acceleration of source (or absorber) that could be scanned to search for the resonance. The red shift problem could be bypassed in a space-borne experiment[16]. The earth's field results in level splitting; it can be canceled statically to a high degree as well as averaged out by swinging the field.

We can now sketch the basic experiment. We envision NbT sources and absorbers (gram scale Nb to ensure *crystal recoil* to hypersharp precision!) made in an identical manner by the TT method of growing He by the decay of T. The resonance signal is the $\tilde{\nu}_e$ induced activity (Rβ). The TT method implies a large T content in the absorber which will create a background (Tβ). A chief design goal is to maximize Rβ/Tβ. The source and absorber are set in the same cryogenic bath at temperatures << 200K. The resonance activation signal, Rβ of 18.6 keV betas grows with time (~t/τ) while the background Tβ decays, thus the rate deviation from the exponential decay (~τ/t) is the resonance signature.

*Table 3 . T-$^3$He hypersharp $\tilde{\nu}_e$ capture rates (worst case, line 1 and best cases lines 2,3 see text). Δt is the delay after start of activation. TT is the "tritium trick".*

| Absorber | Base line | T | $^3$He | Rβ/d | Tβ/d |
|---|---|---|---|---|---|
| TT | 1 cm | 1kCi | 1 pg | $2\times10^6$ (Δt=100d) | $3.7\times10^9$ |
| T-desorbed | 1cm | 1mCi | 1 pg | 2 (Δt=0 d) | ~0 |
| T-desorbed | 10m | 1kCi | 1 pg | 2 (Δt=0 d) | ~0 |

Table 3 shows signal rates in a primitive longitudinal geometry of source-absorber. The 1kCi source is state of the art[17]. The He absorber is made by the TT method and uses only a 1μCi T source to grow 1pg of He in 100 days. The $\tilde{\nu}_e$ activation signal rate Rβ ∝ t (a signature of $\tilde{\nu}_e$ activation) after an initial delay of 100 days, the signal Rβ is 20 Hz vs. the background Tβ of 37 kHz. The signal to background ratio S/B of ~1/2000 may be almost sufficient (with time) to confirm the signal growth vs. the T-decay. However, the S/B can be enhanced as in Table 3 by reducing the T activity by T-desorption via H exchange (by a factor up to ~$10^6$, considered practical[18]). Since the T and He are isolated in trap wells they are relatively isolated from random inhomogenities in the lattice outside the wells due to the exchanged H.

The linear rise of $\tilde{\nu}_e$ activation signal with time assumes a physical age t of the tritium source material comparable to the lifetime; t > 17.8 yr (= τ). Younger T material (t<τ) involves *time-filtered* $\tilde{\nu}_e$ with emission times << τ after T production. This results in a resonance line broadened by the ratio (τ/ t; t<<τ) since the linewidth in this case depends *not on the lifetime τ but on the age t of the source*[19] (which follows simply from time-energy uncertainty). As t→τ, the line becomes narrower, thus the signal at zero detuning rises as ~$t^2$ (t<< τ) instead of as t. This effect leads directly to the prevailing *fundamental* level width regardless of the sources of line broadening from the T/He environment.

Mead[20] has suggested that the Planck length $\mathcal{L}$ in nature would limit the minimum widths of nuclear states rather than the quantum time-energy uncertainty. The $(\Delta E/E)_{min}$ could depend on $\mathcal{L}$ via the imprecision it imposes on r in the nuclear potential V(r). With $\mathcal{L}= (G\hbar/c^3)^{1/2} \sim 10^{-33}$ cm, Mead predicts $\Delta E/E(\mathcal{L}) = (\mathcal{L}/R)\,\beta \sim 10^{-20}$ for β = 1 to ~$10^{-40}$ for $\beta=(\mathcal{L}/R)$ (R is the nuclear radius). The form of β depends on the quantum gravity model. ΔE/E~$10^{-29}$ from T $\tilde{\nu}_e$ capture is ideal to probe this prediction. Indeed, ΔE/E~$10^{-20}$ implies an easily detectable Planck "broadening" by a factor $10^9$.

The Planck broadening effect can be confirmed specifically and directly via the time-filtered resonance effect[19] with tritium sources of different physical age, or the $t^2$ dependence of the signal rate as described above. The effect directly calibrates the signal rate to a quantum resonance width controllable simply by the physical age of the T material. The T age dependence leads to the observable limiting width $\Gamma_{exp}$ of the level. A result $\Gamma_{exp} > \Gamma_\tau$ inferred from the measured lifetime τ of $^3$H should directly expose a fundamental breakdown of time-energy uncertainty in the regime of extremely small energies ~$10^{-24}$ eV. With ref. 20, this result can be interpreted as laboratory evidence for the Planck length.

I wish to thank Don Cowgill and his colleagues at Sandia Labs, Richard Kadell (at LBNL) and Lay Nam Chang, Djordje Minic and Kyunghwa Park (at VT).

[1] J. N. Bahcall, Phys. Rev. **124** (1961) 495
[2] M.Visscher Phys. Rev. **116** (1959) 1581
[3] W. Kells and J .Schiffer, Phys. Rev. **C28** (1983) 2162
[4] R. S. Raghavan, ArXiv: *hep-ph/0601079 (2006)*. The validity of expecting $\tilde{\nu}_e$ flavor oscillations in this experiment has been stressed in A. G. Cohen, S. Glashow and Z. Ligeti, *hep-ph 0810.4602 (2008)* and E. Akhmedov et al, JHEP **0803** (2008) 005.
[5] C. P. Slichter, *Principles of Magnetic Resonance,* (Harper & Row) (1963) p.154
[6] Y. A. Ilinsky and V. Namiot, Sov. J. Quant. Electron. **4** (1975) 891
[7] M. Salkola & S. Stenholm Phys. Rev. **A41** (1990) 3838
[8] F. L. Shapiro, Sov. Phys Uspekhi, **4** (1961) 883
[9] H. Frauenfelder, *The Mőssbauer Effect* (Benjamin)(1962) p19
[10] R. Lässer, *Tritium and $^3$He in Metals* (Springer) (1989)
[11] D. F. Cowgill, Sandia Natl. Lab. Report 2004-1739
[12] M. J. Puska R. M. Nieminen, Phys. Rev. **B29** (1984) 5382
[13] S. T. Picraux, Nucl. Inst. Meth. **182-183** (1981) 413
[14] H. J. Lipkin, Annals of Phys. **9** (1960) 332
[15] J. J. Rush et al, Phys. Rev. **B24** (1981) 4903
[16] S. Glashow (Private Communication)
[17] G. C. Abell & S. Attalla, Phys. Rev. Lett. **59** (1987) 995
[18] Demonstrated in G.C. Abell & D.F. Cowgill, Phys. Rev. **B44** (1991)4178; C. Weinheimer (Priv. Comm.)
[19] C. S. Wu et al, Phys. Rev Lett. **5**, (1960) 432; F. J. Lynch et al, Phys. Rev. **120** (1960) 513; E. Akhmedov et al (in ref. 4) pointed out the time filtering effect in tritium $\tilde{\nu}_e$ experiment.
[20] C. A. Mead, Phys. Rev. **143** (1965) 990